\documentclass[aps,print,superscriptaddress]{revtex4-1}

\usepackage{amsmath,amssymb,textcomp,calc,capt-of,ifthen}
\usepackage{times}
\usepackage{graphicx} 
\usepackage[english]{babel}
\newcommand{\suppMat}{\cite{SupplementaryMaterial}}

\begin{document}

\title{Phonon-mediated quantum state transfer and remote qubit entanglement}

\author{A. Bienfait}
\affiliation{Institute for Molecular Engineering, University of Chicago, Chicago IL 60637, USA}
\author{K. J. Satzinger}
\altaffiliation[Present address: ]{Google, Santa Barbara CA 93117, USA.}
\affiliation{Department of Physics, University of California, Santa Barbara CA 93106, USA}
\affiliation{Institute for Molecular Engineering, University of Chicago, Chicago IL 60637, USA}
\author{Y. P. Zhong}
\affiliation{Institute for Molecular Engineering, University of Chicago, Chicago IL 60637, USA}
\author{H.-S. Chang}
\affiliation{Institute for Molecular Engineering, University of Chicago, Chicago IL 60637, USA}
\author{M.-H. Chou}
\affiliation{Institute for Molecular Engineering, University of Chicago, Chicago IL 60637, USA}
\affiliation{Department of Physics, University of Chicago, Chicago IL 60637, USA}
\author{C. R. Conner}
\affiliation{Institute for Molecular Engineering, University of Chicago, Chicago IL 60637, USA}
\author{\'E . Dumur}
\affiliation{Institute for Molecular Engineering, University of Chicago, Chicago IL 60637, USA}
\affiliation{Institute for Molecular Engineering and Materials Science Division, Argonne National Laboratory, Argonne IL 60439, USA}
\author{J. Grebel}
\affiliation{Institute for Molecular Engineering, University of Chicago, Chicago IL 60637, USA}
\author{G. A. Peairs}
\affiliation{Department of Physics, University of California, Santa Barbara CA 93106, USA}
\affiliation{Institute for Molecular Engineering, University of Chicago, Chicago IL 60637, USA}
\author{R. G. Povey}
\affiliation{Institute for Molecular Engineering, University of Chicago, Chicago IL 60637, USA}
\affiliation{Department of Physics, University of Chicago, Chicago IL 60637, USA}
\author{A. N. Cleland}
\affiliation{Institute for Molecular Engineering, University of Chicago, Chicago IL 60637, USA}
\affiliation{Institute for Molecular Engineering and Materials Science Division, Argonne National Laboratory, Argonne IL 60439, USA}


\date{\today}

\begin{abstract}
Phonons, and in particular surface acoustic wave phonons, have been proposed as a means to coherently couple distant solid-state quantum systems. Recent experiments have shown that superconducting qubits can control and detect individual phonons in a resonant structure, enabling the coherent generation and measurement of complex stationary phonon states. Here, we report the deterministic emission and capture of itinerant surface acoustic wave phonons, enabling the quantum entanglement of two superconducting qubits. Using a 2 mm-long acoustic quantum communication channel, equivalent to a 500 ns delay line, we demonstrate the emission and re-capture of a phonon by one qubit; quantum state transfer between two qubits with a 67\% efficiency; and, by partial transfer of a phonon between two qubits, generation of an entangled Bell pair with a fidelity of $\mathcal{F}_B = 84 \pm 1$\%.
\end{abstract}


\maketitle


Electromagnetic waves, whether at optical or at microwave frequencies, have to date played a singular role as carriers of quantum information between distant quantum nodes, providing the principal bus for distributed quantum information processing. Several recent experiments have used microwave photons to demonstrate deterministic as well as probabilistic remote entanglement generation between superconducting qubits, with entanglement fidelities in the range of 60\%-95\% \cite{kurpiersDeterministicQuantumState2018, campagne-ibarcqDeterministicRemoteEntanglement2018, axlineOndemandQuantumState2018, leungDeterministicBidirectionalCommunication2018, zhongViolatingBellInequality2018}. However, for some solid-state quantum systems, such as electrostatically-defined quantum dots or electronic spins, the strong interactions with the host material make acoustic vibrations, or phonons, an alternative and potentially superior candidate to photons. In particular, surface-acoustic wave (SAW) \cite{morgandavidSurfaceAcousticWave2007} phonons have been proposed as a universal medium for coupling remote quantum systems \cite{barnesQuantumComputationUsing2000, schuetzUniversalQuantumTransducers2015}. These also have the potential for efficient conversion between microwave and optical frequencies \cite{vainsencherBidirectionalConversionMicrowave2016, shumeikoQuantumAcoustoopticTransducer2016}, linking microwave qubits to optical photons. These proposals followed pioneering experiments showing the coherent emission and detection of traveling SAW phonons by a superconducting qubit  \cite{gustafssonLocalProbingPropagating2012,gustafssonPropagatingPhononsCoupled2014}. Traveling SAW phonons have also been used to transfer electrons between quantum dots \cite{hermelinElectronsSurfingSound2011, mcneilOndemandSingleelectronTransfer2011}, couple to nitrogen-vacancy centers \cite{golterCouplingSurfaceAcoustic2016} as well as drive silicon carbide spins \cite{whiteleyProbingSpinphononInteractions2018}. When localized in Fabry-P{\'e}rot resonators, standing-wave SAW phonons have been coherently coupled to superconducting qubits \cite{manentiCircuitQuantumAcoustodynamics2017, mooresCavityQuantumAcoustic2018, bolgarQuantumRegimeTwoDimensional2018,
noguchiSinglephotonQuantumRegime2018, satzingerQuantumControlSurface2018}, allowing the on-demand creation, detection and control of quantum acoustic states \cite{satzingerQuantumControlSurface2018}. Related experiments have demonstrated similar control of localized bulk acoustic phonons \cite{oconnellQuantumGroundState2010, chuQuantumAcousticsSuperconducting2017, chuCreationControlMultiphonon2018}.

Here, we report the use of itinerant SAW phonons to realize the coherent transfer of quantum states between two superconducting qubits. We first show that a single superconducting qubit can launch an itinerant phonon into a SAW resonator when operating near the strong multi-mode coupling regime where the coupling between the qubit and one Fabry-P{\'e}rot mode exceeds the resonator free spectral range. This allows the phonon to be completely injected into the acoustic channel before any re-excitation of the emitting qubit. Using techniques developed for microwave photon transfer \cite{korotkovFlyingMicrowaveQubits2011, zhongViolatingBellInequality2018}, we show that the emitting qubit can re-capture the phonon at a later time, with a 67\% efficiency. Using the same acoustic channel, we also perform two-qubit quantum state transfer as well as remote qubit entanglement with a fidelity exceeding 80\%.


The experimental layout is shown in Fig. 1. The acoustic part of the device is a SAW resonator, with an effective Fabry-P{\'e}rot mirror spacing of 2 mm, corresponding to a single-pass itinerant  phonon travel time of about 0.5~$\mu$s.  The resonator is coupled to two frequency-tunable superconducting Xmon qubits \cite{barendsCoherentJosephsonQubit2013}, $Q_1$ and $Q_2$; their coupling is electrically controlled using two tunable couplers, $G_1$ and $G_2$ \cite{chenQubitArchitectureHigh2014}. The tunable couplers, the qubits, and their respective control and readout lines are fabricated on a sapphire substrate, while the SAW resonator is fabricated on a separate lithium niobate substrate. The qubits have native relaxation times $T_{1,\mathrm{int}}$ of 22 and 26 $\mu$s and coherence times $T_{2,\mathrm{Ramsey}}$ of 2.1 and 0.6~$\mu$s.

The SAW resonator is defined by two acoustic mirrors, which are arrays of 30-nm-thick aluminum lines, spaced by 0.5~$\mu$m, with 400 lines in each array, defining two Bragg mirrors on each side of the central acoustic emitter-receiver. The acoustic emitter is an interdigitated transducer (IDT), comprising forty 30-nm thick lines with alternate lines connected to a common electrical port. An electrical pulse applied to the IDT results in two symmetric SAW pulses traveling in opposite directions, reflecting off the mirrors, and completing a round trip in a time $\tau = 508$~ns \suppMat. The mirrors support a 125 MHz wide stop-band, centered at 3.97 GHz, localizing about $60$ Fabry-P{\'e}rot standing modes with a free-spectral range  $\nu_{\mathrm{FSR}} = 1/\tau = 1.97$~MHz. Each port of the IDT is inductively grounded, each inductor forming a mutual inductance through free space with a similarly-grounded inductance in the couplers $G_1$ and $G_2$ on the sapphire chip. The coupled inductors are precisely aligned to one another in a flip-chip assembly \cite{satzingerQuantumControlSurface2018, satzingerSimpleNongalvanicFlipchip2018}. Each coupler comprises a $\pi$ inductive network, with a Josephson junction bridging two fixed inductors to ground, one of the fixed inductors coupled to one port of the SAW transducer and the other contributing to the qubit inductance. A flux line controls the phase $\delta_{\mathrm{i}}$ across the coupling Josephson junction, thus controlling its effective inductance and the qubit-IDT coupling. Each coupler can be switched from maximum coupling to off in a few nanoseconds \cite{chenQubitArchitectureHigh2014}, isolating the qubits \cite{satzingerQuantumControlSurface2018}.

The qubits' controlled coupling to the IDT enables the time domain-shaped emission of itinerant phonons into the resonator \cite{satzingerQuantumControlSurface2018}. We characterize this emission by exciting the qubit and then monitoring its excited state population, shown in Fig. 1e, with the excited state decaying due to phonon emission. As expected, the frequency dependence of the decay rate follows closely the frequency response of the IDT's electrical admittance, see \cite{satzingerQuantumControlSurface2018, SupplementaryMaterial}. When near maximum admittance, the qubit relaxes in $T_1 < 15$ ns, a value roughly 3\% of the phonon's resonator transit time $\tau$ and less than 0.1\% of the qubit's intrinsic relaxation time $T_{1,\mathrm{int}}$. The relaxation rate of the qubit also sets the extent of the emitted phonon wavepacket: the phonon can truly be considered as an itinerant excitation, with each oppositely-traveling packet having a spatial extent of about 60~$\mu$m, compared to the 2~mm length of the resonator. For phonon frequencies within the mirror stop-band, the emitted itinerant phonon is strongly reflected by the mirrors, and can re-excite the qubit at integer multiples of the phonon transit time $t = n\tau$ (Fig. 1e). We note that the transit time is unusually long for superconducting qubit experiments: the equivalent microwave cable length would be $\sim 100$ m, much larger than the typical $\sim 1$ m connections used for quantum communication experiments \cite{kurpiersDeterministicQuantumState2018, campagne-ibarcqDeterministicRemoteEntanglement2018, axlineOndemandQuantumState2018, leungDeterministicBidirectionalCommunication2018, zhongViolatingBellInequality2018}. In the rest of this work, we fix the qubits' operating frequencies to $\omega_{\mathrm{Q,i}}/2\pi$ = 3.95 GHz, well within the mirror stop-band.

\begin{figure}[t]
\includegraphics[width=8.6cm]{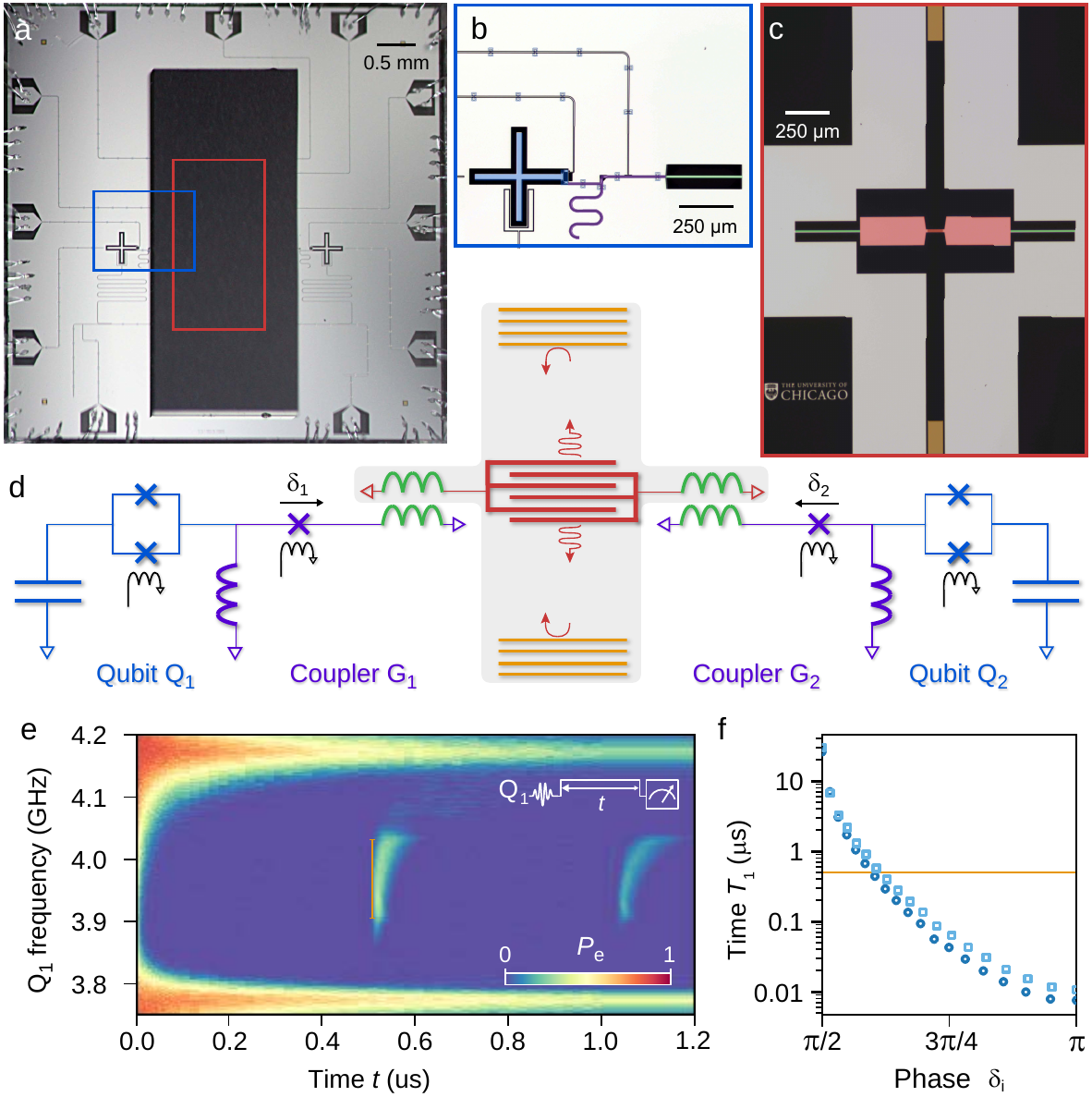}
\centering
\caption{Experimental device. (a)  Micrograph of flip-chip assembled device, with (b) two superconducting qubits ($Q_1$ and $Q_2$, blue), connected to two tunable couplers ($G_1$ and $G_2$, purple), fabricated on sapphire. These are connected via two overlaid inductors (green) to (c) a SAW resonator, fabricated on lithium niobate. The SAW resonator comprises two Bragg mirrors (orange), spaced by 2 mm, defining a Fabry-P{\'e}rot acoustic cavity probed by an interdigitated transducer (red). (d) Simplified circuit diagram, the grey box indicating elements on the flipped lithium niobate chip. (e) Excited state population $P_{\mathrm{e}}$ for qubit $Q_1$, with coupler $G_1$ set to maximum and $G_2$ turned off. $Q_1$ is prepared in $\vline e \rangle$ using a $\pi$ pulse, its frequency set to $\omega_{\mathrm{Q}1}$ (vertical scale) for a time $t$ (horizontal scale), before dispersive readout of its excited population $P_{\mathrm{e}}$ \suppMat. $Q_1$ relaxes due to phonon emission via the IDT, and if its frequency is within the mirror stop-band from 3.91 to 4.03 GHz, the emitted phonon is reflected and generates qubit excitation revivals at times $\tau$ (orange line) and $2\tau$. Inset: Pulse sequence. (f) Measured qubit energy decay time $T_1$ for $\omega_{\mathrm{Q,i}}/2\pi = 3.95$~GHz as a function of the coupler Josephson junction phase $\delta_{\mathrm{i}}$, showing the qubit emission can be considerably faster than the phonon transit time (orange line), for both $Q_1$ (circles) and $Q_2$ (squares).}
\end{figure}

We first explore the emission and re-capture of phonons by one qubit. The highest efficiencies are  achieved by controlling the qubit-acoustic channel coupling rates $\kappa_i$ so as to match the phonon envelope during the emission and capture of the phonon \cite{korotkovFlyingMicrowaveQubits2011, campagne-ibarcqDeterministicRemoteEntanglement2018, kurpiersDeterministicQuantumState2018, axlineOndemandQuantumState2018, zhongViolatingBellInequality2018}. Each qubit's coupling rate $\kappa_i$ is controlled by the coupler $G_{\mathrm{i}}$, and is calibrated by measuring the qubit energy decay rate $T_1 ^{-1}= \kappa_i+T_{1,\mathrm{int}}^{-1}$ as a function of the coupler flux bias, then subtracting the contribution from the intrinsic qubit lifetime $T_{1,\mathrm{int}}$ (see Fig. 1f). At maximum coupling, we measure $1/\kappa_1 = 7.6$ ns and $1/\kappa_2 = 10.6 $ ns, the difference being due to a 5\% mismatch in the couplers' Josephson junction inductances.  At this maximum point, the qubit coupling $g_i$ to an \emph{individual} Fabry-P{\'e}rot mode are $g_1/2\pi = 2.57 \pm 0.1$ MHz and $g_2/2\pi = 2.16 \pm 0.1$ MHz \suppMat. The system is thus at the threshold of ``strong'' multi-mode coupling \cite{SupplementaryMaterial, mooresCavityQuantumAcoustic2018}, where the coupling rate $g_i$ is larger than the resonator free-spectral range, the qubit linewidth, and the SAW energy decay rate (respectively, $\nu_{\mathrm{FSR}}= 1.97$~MHz, $1/T_{2,\mathrm{Ramsey}} = 2 \pi \times 76$~kHz, and $1/T_{1, \mathrm{SAW}} \sim 2 \pi \times 133$~kHz \suppMat).

In Figure 2, we demonstrate a one-qubit single-phonon ``ping-pong'' experiment using qubit $Q_1$. With $G_2$ off, we apply calibrated control pulses to $G_1$ and $Q_1$ to obtain a symmetric phonon envelope of characteristic bandwidth $1/\kappa_c = 10$ ns, and monitor $Q_1$'s excited-state population $P_{\mathrm{e}}$ (Fig. 2a). The emission takes about $150$ ns, after which $P_{\mathrm{e}}$ remains near zero during the phonon transit. After $\sim 0.5$ $\mu$s, the phonon returns and is recaptured, with $P_{\mathrm{e}}$ increasing and leveling off for times $t>t_f = 0.65 \mu$s. The capture efficiency is $\eta = P_{\mathrm{e}}(t_{\mathrm{f}})/P_{\mathrm{e}}(t=0) = 0.67 \pm 0.01$.

Successive transits show a geometric decrease in capture efficiency, $\eta_n = \eta^{n}$ with transit number $n$ (Fig. 2b). We attribute the decrease to losses in the acoustic channel. An independent measurement yields a Fabry-P{\'e}rot mode energy decay rate $T_{1,\mathrm{SAW}} = 1.2~\mu$s \suppMat, close to similar resonators in the literature \cite{manentiCircuitQuantumAcoustodynamics2017, mooresCavityQuantumAcoustic2018}. This sets an upper limit $\eta \lesssim e^{-\tau/T_{1,\mathrm{SAW}}} \approx 0.65$ \suppMat, close to our measurement.

We perform quantum process tomography of the one-qubit release-and-catch operation by preparing four independent qubit states and reconstructing the process matrix $\chi$ at time $t_{\mathrm{f}}$ (Fig. 2c), yielding a process fidelity $\mathcal{F}_1 = \mathrm{Tr}(\chi \cdot \chi_{\mathrm{ideal}}) = 0.83 \pm 0.002$. A master equation simulation, taking into account an acoustic channel loss of $\eta = 0.67$ and the finite qubit coherence $T_{2R} = 2.1~\mu$s, yields a process matrix whose trace distance to $\chi $ is $\sqrt{\mathrm{Tr} \left( \left[ \chi - \chi_{\mathrm{sim}} \right]^2 \right)}$ = 0.07.

\begin{figure}[t]
\centering
\includegraphics[width=8.6cm]{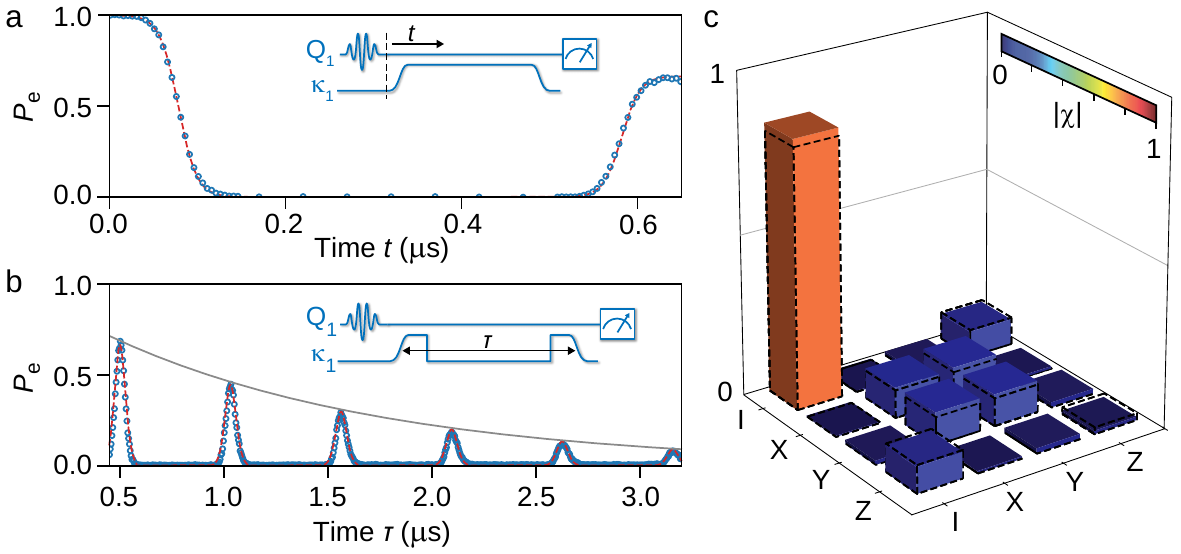}
\caption{Emission and capture of a shaped itinerant phonon. (a) Circles represents the measured excited state population of $Q_1$ when interrupting the sequence after a time $t$. Inset: Control pulse sequence. (b) Measured excited state population of $Q_1$ while sweeping the delay between the emission and capture control pulses, evidencing a population geometrically decreasing with the number of transits (grey line). (c) Quantum process tomography at the maximum efficiency point of (b), with a process fidelity $\mathcal{F}_1 = 0.83 \pm 0.002$. In all panels, dashed lines indicate the results of a master equation simulation including a finite transfer efficiency and qubit imperfections, see\suppMat .}
\centering
\end{figure}

In Figure 3, we demonstrate the interferometric nature of the one-qubit phonon emission and capture. We prepare $Q_1$ in $\lvert e \rangle$, then emit a half-phonon, and capture it with $Q_1$ after one transit. As capture is the time-reversal of emission, it would seem that the photonic half-excitation left in $Q_1$ would be emitted during the capture process. In fact, depending on the relative phase $\Delta \phi$ between the stored half-photon and the reflected  half-phonon, the two half-quanta will interfere destructively or constructively, giving  results ranging from re-excitation of the qubit to total emission. When $\Delta \phi = \pi$, the reflected half-phonon interferes constructively with the emitted half-photon stored in $Q_1$, and all the energy is transferred to the SAW resonator, but when $\Delta \phi = 0$, destructive interference results in qubit re-excitation. Figure 3a shows the qubit population for $\Delta \phi = 0$ and $\Delta \phi = \pi$, with the final population $P_{\mathrm{e}}(t_{\mathrm{f}})$ reaching respectively 0.77 and 0.08. Figure 3b shows that the final qubit population oscillates as expected between these two limiting values as a function of $\Delta \phi$. A simulation including channel loss and qubit dephasing accounts partially for the reduction in interferometric amplitudes, the remaining mismatch being attributed to control pulse imperfections \suppMat.

\begin{figure}[t]
\centering
\includegraphics[width=5.5cm]{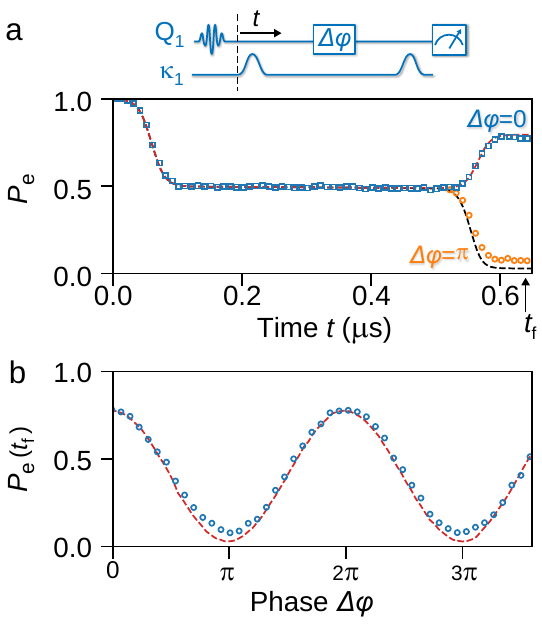}
\caption{Phonon interferometry. With $Q_1$ initially prepared in $|e\rangle$, a control signal on $G_1$ releases and subsequently re-captures half a phonon to the resonator. Simultaneously, a 20 MHz detuning pulse of varying duration is applied to $Q_1$ to change its phase by $\Delta \phi$. (a) Measured $Q_1$ excited state population when interrupting the sequence after a time $t$, with a phase difference $\Delta \phi = 0$ (squares) or $\pi$ (circles). Inset shows the control sequence. (b) $Q_1$ final state $P_e(t = t_{\mathrm{f}})$ for $t_{\mathrm{f}} =0.65~\mu$s as a function of the phase difference $\Delta \phi$ between the half-photon and and half-phonon. Circles are experimental points. Dashed lines are simulations based on an input-output theory model, see \suppMat .}
\centering
\end{figure}

We can use the acoustic communication channel to transfer quantum states, as well as generate remote entanglement between the two qubits. In Fig. 4a, we demonstrate a quantum swap between $Q_1$ and $Q_2$. As the phonon transit time is significantly larger than the phonon temporal extent, up to three itinerant phonons can be stored sequentially in the SAW resonator. We thus can perform a double swap, by having each qubit successively emit an excitation, then at a later time capture the phonon emitted by the other qubit. This process has a fidelity $\mathcal{F}_2 = 0.63 \pm 0.01$ \suppMat, in agreement with the one-qubit state transfer fidelity $\mathcal{F}_2 \sim \mathcal{F}_1^2$ and with infidelity dominated by acoustic losses.

We also use the acoustic channel to share half of an itinerant phonon, generating remote entanglement between $Q_1$ and $Q_2$. We first prepare $Q_1$ in $\lvert e\rangle$, then apply a calibrated pulse to coupler $G_1$ to release half of $Q_1$'s excitation to the acoustic channel. This half-phonon is then captured, after one transit, by $Q_2$, using the pulse sequence shown in Fig. 4b, creating a Bell state $| \Psi \rangle = (| eg \rangle + e^{i\alpha} | ge \rangle) / \sqrt{2}$, the phase $\alpha$ resulting from the relative de-tunings of the qubits during the sequence. The excited state population of each qubit is shown as a function of time in Fig. 4b, and the Pauli matrices and reconstructed density matrix $\rho$, at time $t=0.65~\mu$s, are shown in Fig. 4c and d. We find a state fidelity, referenced to the ideal Bell state, of {$\mathcal{F}_B = \mathrm{Tr}(\rho \cdot \rho_{| \Psi \rangle}) = 0.84 \pm 0.01$ and a concurrence $C = 0.61 \pm 0.04$. A master equation simulation yields a density matrix $\rho_{\mathrm{sim}}$, which has a small trace distance $\sqrt{\mathrm{Tr} \left( \left[ \rho - \rho_{\mathrm{sim}} \right]^2 \right)} = 0.06$ to $\rho$, with errors dominated by acoustic losses.

\begin{figure}[t]
\centering
\includegraphics[width=8.6cm]{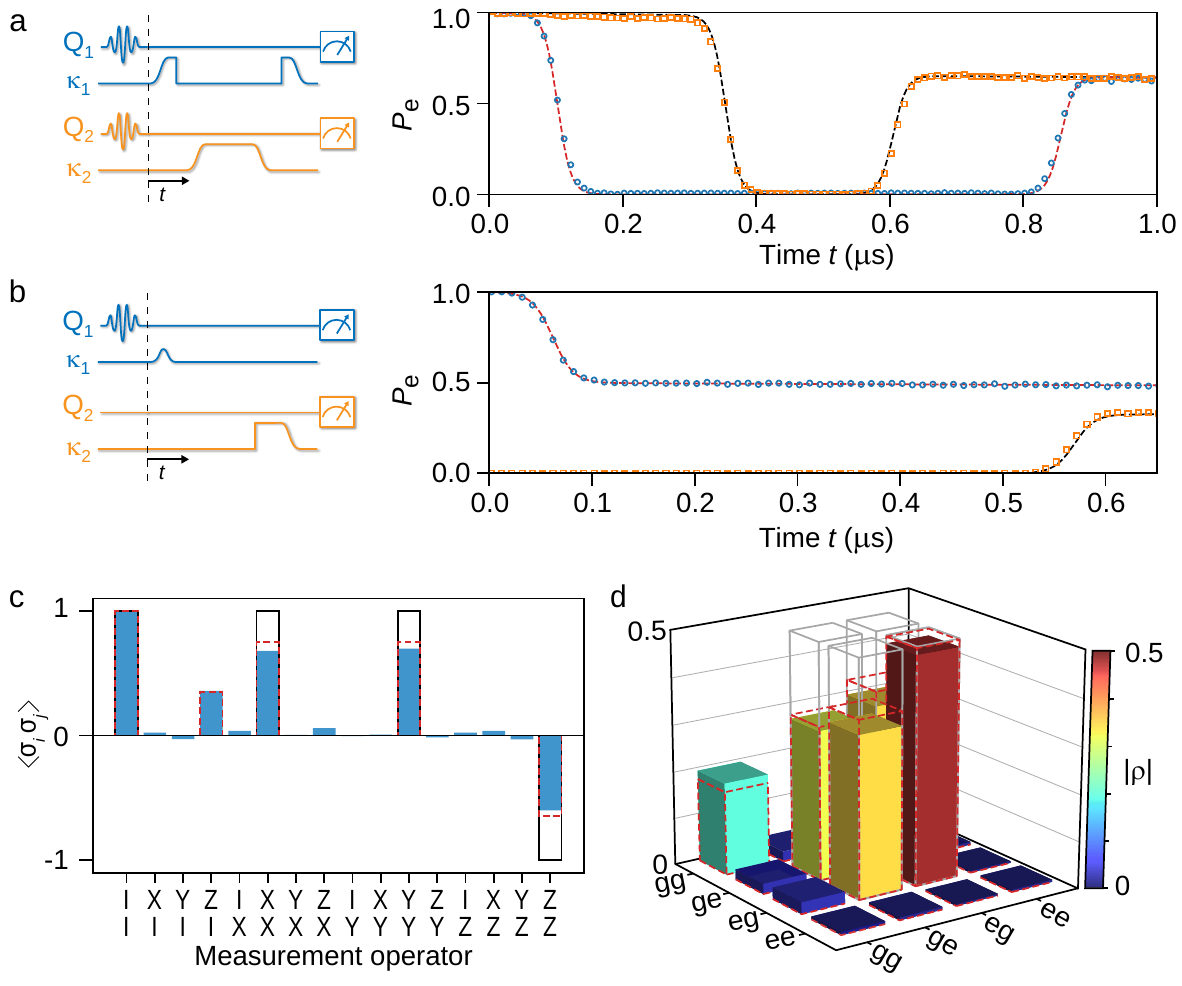}
\caption{Quantum state transfer and remote entanglement. (a) Qubit state swap via the acoustic channel, with control pulses shown in left panel. (b) Acoustic entanglement: With $Q_1$ initially in $| e \rangle$, a control signal applied to $G_1$ releases half a phonon to the channel, captured later by $Q_2$.  In (a) and (b), circles and squares are $Q_1$ and $Q_2$ excited state populations measured simultaneously after a time $t$. (c) Expectation values of two-qubit Pauli operators for the reconstructed Bell state density matrix (d) at $t=0.65~\mu$s.  In (c) and (d), solid lines indicate values expected for the ideal Bell state $| \Psi \rangle = (| eg \rangle + | ge \rangle) / \sqrt{2}$. In all panels, dashed lines are simulation results including a finite transfer efficiency and qubit imperfections \suppMat.}\centering
\end{figure}

These results comprise clear and compelling demonstrations of the controlled release and capture of itinerant phonons into a confined Fabry-P{\'e}rot resonator, and are limited primarily by acoustic losses. As the phonons have a spatial extent much smaller than the resonator length, the emission and capture processes are ``blind'' to the length of the resonator, so the same processes would work in a non-resonant acoustic device. We demonstrate that these processes can generate quite high fidelity entanglement between two qubits. These results constitute an important step towards realizing fundamental quantum communication protocols with phonons. Further demonstrations could include the use of directional transducers \cite{collinsUnidirectionalSurfaceWave1969, ekstromSurfaceAcousticWave2017}, and lower-loss acoustic materials. This experiment also opens the door to future tests of entanglement using surface acoustic wave phonons: A first step could be to probe the entanglement of the counter-propagating wavepackets emitted by the IDT in the resonator, further tests could be demonstrations of violations of the Leggett-Garg  \cite{leggettQuantumMechanicsMacroscopic1985} or Bell inequalities \cite{bellEinsteinPodolskyRosen1964}, as well as exploring delayed-interaction phenomena\cite{guoGiantAcousticAtom2017}.

\begin{acknowledgments}
We thank P. J. Duda for helpful discussions and we thank W.D. Oliver and G. Calusine at Lincoln Laboratories for the provision of a traveling-wave parametric amplifier (TWPA). Devices and experiments were supported by the Air Force Office of Scientific Research and the Army Research Laboratory. K.J.S. was supported by NSF GRFP (NSF DGE-1144085),
\'E.D. was supported by LDRD funds from Argonne National Laboratory, and A.N.C. was supported by the DOE, Office of Basic Energy Sciences. This work was partially supported by the UChicago MRSEC (NSF DMR-1420709) and made use of the Pritzker Nanofabrication Facility, which receives support from SHyNE, a node of the National Science Foundation's National Nanotechnology Coordinated Infrastructure (NSF NNCI-1542205). The authors declare no competing financial interests. The datasets supporting this work are available from the corresponding author on request. Correspondence and requests for materials should be addressed to A. N. Cleland (anc@uchicago.edu). 
\end{acknowledgments}

\clearpage

\begin{center}
\textbf{\Large{Supplementary Materials for Phonon-mediated quantum state transfer and remote qubit entanglement}}
\end{center}

\setcounter{figure}{0}   
\renewcommand{\thefigure}{S\arabic{figure}}
\renewcommand{\thetable}{S\arabic{table}}
\renewcommand{\theequation}{S\arabic{equation}}

\section{Device, experimental setup and techniques}

The experiment is conducted using a dilution refrigerator with a base temperature below 10 mK. A full wiring diagram and a description of the room-temperature set-up may be found in Ref. \cite{zhongViolatingBellInequality2018}. The qubits and the SAW resonator used here have related designs to, and were fabricated following the same process as, those described in Ref. \cite{satzingerQuantumControlSurface2018}. The major difference is that here the SAW resonator mirrors are placed 2~mm apart, rather than a few microns, giving a much larger travel time for phonons released into the resonator.

The device parameters are given in Table S1. The sapphire and lithium niobate chips are fabricated separately, after which they are assembled: One of the chips is flipped upside down and aligned to the other chip using a contact mask aligner. The two chips are held together with a lateral precision of better than 1 $\mu$m, and are vertically spaced by $\sim 5~\mu$m. The photoresist used to hold them together is applied before this process and dried as described in Ref. \cite{satzingerQuantumControlSurface2018}.

\renewcommand{\baselinestretch}{1.}
\begin{table}[h]
\center
  \begin{tabular}{p{9cm}ccc}
    \hline
    \hline
    Qubit parameters & \  & Qubit 1 & Qubit 2\\
    \hline
    Qubit bare frequency (GHz) & \  & 4.84 & $\sim 6$\\
    Qubit capacitance (design value) (fF) & \  & 90 & 90\\
    SQUID inductance (nH) & \  & 11.2 & 7.5\\
    Qubit anharmonicity (MHz) & \  & 179 & 188\\
    Qubit intrinsic lifetime, $T_{1, \mathrm{int}}$ ($\mu$s) & \
    & 21.7(1) & 26.1(1)\\
    Qubit Ramsey dephasing time, $T_{2,\mathrm{Ramsey}}$ ($\mu$s) & \  &
    2.10(4) & 0.60(1)\\
    Qubit spin-echo dephasing time, $T_{2 E}$ ($\mu$s) & \  &
    6.29(2) & 1.84(1)\\
    $| e \rangle$ state readout fidelity & \  & 0.933(8) & 0.952(8)\\
    $| g \rangle$ state readout fidelity & \  & 0.969(6) & 0.977(5)\\
    Thermal excited state population, without acoustic cooling & \  &
    0.0281(5) & 0.0223(5)\\
    Thermal excited state population, with acoustic cooling & \  & 0.0059(4) &
    0.0053(3)\\
    Readout resonator frequency (GHz) & \  & 5.251 & 5.293\\
    Readout resonator quality factor & \  & $3.5 \times 10^3$ & $1.4 \times
    10^3$\\
    Readout dispersive shift (MHz) & \  & 1.5 & 1.1\\
    \hline \hline
    \  & \  & \  & \ \\
    \hline \hline
    Tunable coupler parameters & \  & Coupler 1 & Coupler 2\\
    \hline
    Coupler junction inductance (nH) & \  & 1.14(1) & 1.18(1)\\
    IDT grounding inductance (design value) (nH) & \  & 0.4 & 0.4\\
    Coupler grounding inductance (design value) (nH) & \  & 0.4 & 0.4\\
    Mutual coupling inductance between IDT and coupler (nH) & \  & 0.21(2) &
    0.21(2)\\
    \hline \hline
    \  & \  & \  & \ \\
    \hline \hline
    SAW resonator parameters & Free space & Mirror  & Transducer\\
    \hline
    Aperture (\textmu m) & \  & 75 & 75\\
    Wave propagation speed (km/s) & 4.034(2) & 3.928(2) & 3.911(2)\\
    Wave propagation losses (Np/m) & 70(10) & - & -\\
    Number of cells & \  & 400 & 20\\
    Pitch (\textmu m) & \  & 0.5 & 0.985\\
    Reflectivity & \  & -0.049$i$(5) & 0.009$i$(2)\\
    Metallization ratio & \  & 0.58 & 0.58\\
    Effective mirror-mirror distance ($\mu$m) &   & 2029.6 &   \\
    Free spectral range (MHz) &  & 1.97 &   \\
    \hline \hline
  \end{tabular}
  \caption{\label{tableone}Device parameters for the two qubits, parameters related to the interdigitated acoustic transducer (IDT), the tunable couplers connecting each qubit to the SAW resonator, and the SAW resonator itself.}
\end{table}
\renewcommand{\baselinestretch}{1.5}

\subsection{Qubits}

Each qubit can be tuned from its maximum (bare) frequency (see Table S1) to frequencies below
3.7 GHz, using a $Z$-control flux bias threaded through the qubit SQUID loop. All  results
presented in the main text are acquired using the qubit operating frequency of
3.95 GHz, with idling and readout frequencies of 4.17 and 4.22 GHz for qubits $Q_1$ and $Q_2$ respectively, close to a node in the interdigitated transducer (IDT) response. This was done to avoid spurious acoustic relaxation. In addition, in the idle state, each coupler
flux is set to minimize the coupling between the qubits and SAW resonator, with an estimated minimum coupling $g/2\pi$ to an individual SAW Fabry-P{\'e}rot mode of less than $70$~kHz. The qubit coherence times $T_{1,\mathrm{int}}$ and $T_2$ in this idle
state are given in Table S1. We note there is very little
variation in these values over the frequency range of interest in our experiment
(3.8 to 4.2 GHz), provided the couplers are set to their minimum coupling values.

The qubits are read out using the induced dispersive frequency shift of each qubit's readout resonator. To
have an optimal readout fidelity, we use a $\lambda / 2$ Purcell filter \cite{jeffreyFastAccurateState2014} common to the
two readout resonators, and we use a traveling-wave amplifier \cite{macklinQuantumlimitedJosephsonTravelingwave2015} (Lincoln Laboratories) followed by a
cryogenic high electron mobility transistor amplifier (Low Noise Factory) to
ensure near quantum-limited amplification. We measure both qubits
simultaneously using a 600 ns-long readout tone, and discriminate between the
four joint basis states $\left \lbrace | gg \rangle, | ge \rangle, | eg \rangle, | ee \rangle \right \rbrace$. The qubits' readout state
fidelities are given in Table S1. In the main text, we show qubit measured populations corrected for these readout errors, while all quantum process and quantum state tomography fidelity calculations are done on non-corrected data. Table S2 shows that there is little difference between the fidelity values obtained with or without readout correction.

We generate controlled Rabi rotations in each qubit using shaped microwave pulses
applied to a dedicated XY microwave drive line capacitively coupled to each qubit. To prevent spurious excitation
of one qubit by the other qubit's XY drive, the idling qubit is de-tuned by $-450$ MHz during
the application of this drive pulse. Using Rabi oscillations combined with joint simultaneous measurements, we measure less than 0.5\% spurious excitation errors.

A similar SAW resonator, see Ref. \cite{satzingerQuantumControlSurface2018}, displayed less than 0.5\% spurious thermal excitation,
which is lower than is typically observed in superconducting qubits. We take
advantage of this feature and use the resonator to cool the qubits before each measurement. This is done by
coupling each qubit in turn for 100 ns to the IDT mode at $3.85$ GHz, outside the mirror bandwidth, sufficiently long for the qubits to relax to the SAW effective temperature by emission of phonons. Using Rabi population measurements \cite{geerlingsDemonstratingDrivenReset2013} we find that after this protocol, the qubits have a spurious excited state population of $\sim 0.6$\%, compared to $\sim 3$\% without the cooling protocol; see Fig.~\ref{rpm}.

\begin{figure}[t]
\centering
\includegraphics[width = 10cm]{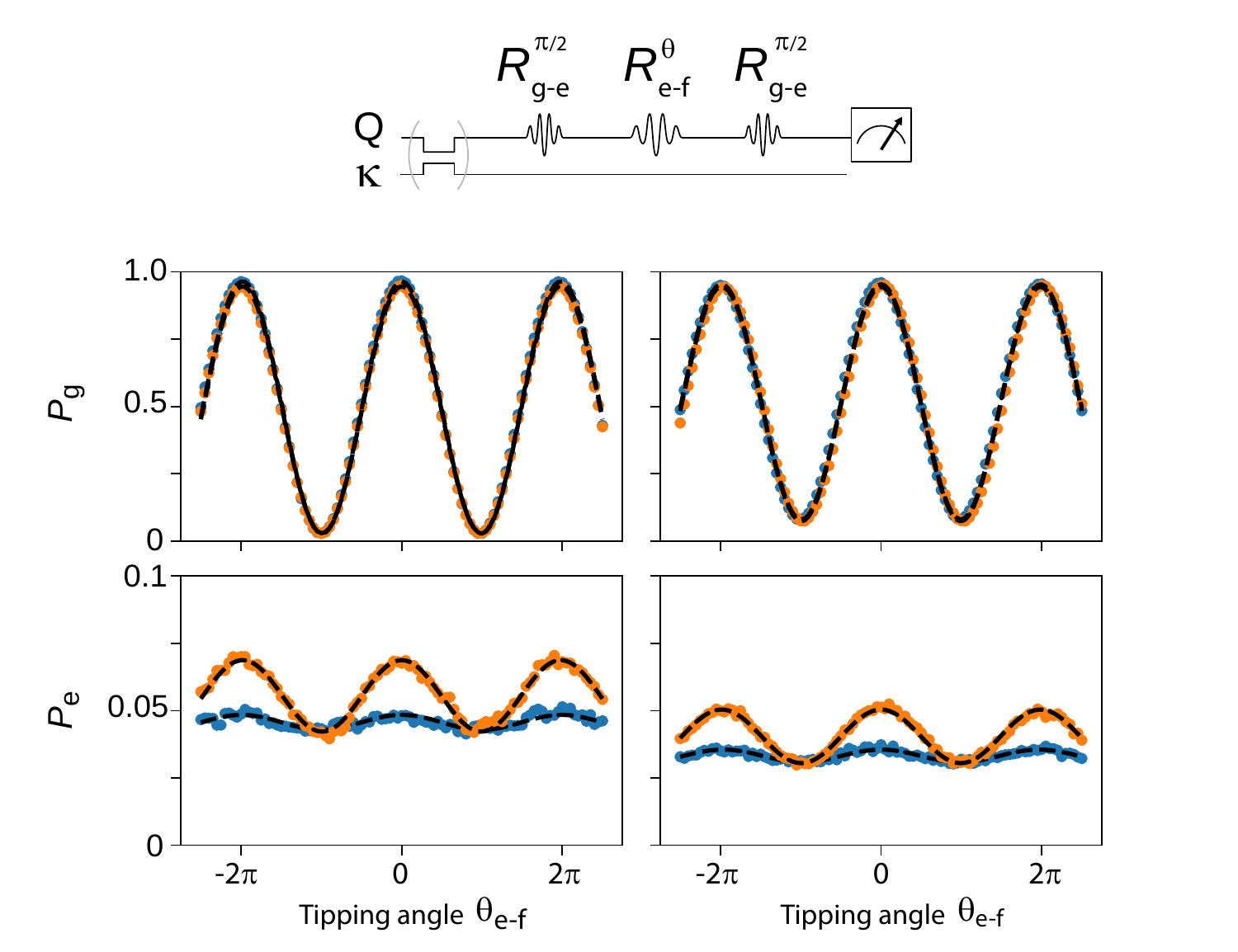}
\caption{\label{rpm}Acoustic cooling and spurious excited state population. Rabi population measurement (control sequence at the top) \cite{geerlingsDemonstratingDrivenReset2013} with (blue points) and without (orange points) acoustic cooling. Lines are cosine fits. The cooling protocol reduces the thermal population by about a factor of five.}
\centering
\end{figure}

We perform quantum state tomography of each qubit by applying the tomography
gates $\left \lbrace I, R_x^{\pi / 2}, R_y^{\pi / 2} \right \rbrace$, following which we perform a joint
dispersive readout as described above. The density matrix is reconstructed via
linear inversion and we ensure it is Hermitian and positive semi-definite with
unit trace.

Quantum process tomography of single and two-qubit state transfers is
performed by preparing the emitter qubit(s) in the input states $\left \lbrace | g
\rangle, (| g \rangle + | e \rangle) / \sqrt{2}, (| g \rangle + i | e \rangle)
/ \sqrt{2}, | e \rangle \right \rbrace$ and then performing the state transfer.
Measurements of the associated density matrices of the receiver (emitter)
qubit(s) are performed after the state transfer (preparation) as described
above, and the process matrix is obtained through linear inversion from these
matrices. We ensure that the process matrix is Hermitian and positive semi-definite.

\subsection{Tunable couplers}\label{Couplers}

The tunable couplers $G_j$ are used to control the interaction between the
qubits and the SAW resonator. The couplers are adjusted by applying an external magnetic flux through a control wire that changes their Josephson junction inductance. Adjusting a coupler also modifies the inductance of the qubit connected to that coupler,  and thus induces a shift of the qubit frequency.

We calibrate this frequency shift experimentally for all qubit operating frequencies and coupler biases. We also measure each qubit's acoustic emission rate for all qubit operating frequencies and coupler biases. This calibration allows us to execute a particular control sequence by specifying the desired qubit operating frequencies and coupling rates $\kappa_i$. We then determine the control pulses needed to apply to the qubit and coupler flux lines, correcting for the resulting frequency shifts.

We also use the coupler-induced qubit frequency shift to calibrate each coupler's
effective Josephson inductance. This measurement is done at a node of the IDT
response (4.17 GHz), where all phonon emission is suppressed. The results of these calibrations are shown in Table S1.

\subsection{SAW resonator}

Electron micrographs of the interdigitated transducer (IDT) and the Bragg mirror fingers are shown in
Fig. \ref{SAW}. The transducer pitch, which is twice the transducer finger-spacing, is chosen to be slightly smaller than twice the mirror finger spacing, so as to better match the transducer's frequency response to the mirrors' stop-band (see Table S1). The SAW
resonator parameters shown in Table S1 are determined as follows:

\begin{itemize}
  \item {\textbf{Mirrors:}} The mirrors' effective wave propagation speed
  and reflection, both parameters in a SAW coupling-of-modes (COM) model \cite{morgandavidSurfaceAcousticWave2007}, are tuned to reproduce the experimental stop-band, shown in Fig. 1e as an orange line.

  \item {\textbf{Transducer:}} We combine a circuit model for the
  qubit and its coupler with a standard one-dimensional electromechanical model, the
  $P$-matrix model \cite{morgandavidSurfaceAcousticWave2007}, to simulate the IDT response. We use the circuit model to
  determine the qubit frequency and decay time \cite{niggBlackBoxSuperconductingCircuit2012} as a function of the qubit's SQUID
  flux bias. The transducer parameters are adjusted to reproduce the qubit
  $Q_1$ decay rates observed in Fig. 1e; see Fig. \ref{SAW}.

  \item {\textbf{Free space:}} The free-space SAW propagation speed is
  determined by matching the observed phonon transit time $\tau$ to the
  resonator geometrical length, taking into account the delays induced by the
  transducer and the mirror, and the effective reflection point in each mirror. A separate measurement of the SAW resonator shows that the SAW Fabry-P{\'e}rot modes have
  intrinsic quality factors typically in the range of $4 \times 10^4 - 6 \times 10^4$,
  yielding an energy relaxation time $T_{1 \mathrm{SAW}} ~ \sim ~ 2$ \textmu s, close to what is experimentally inferred for this sample ($T_{1 \mathrm{SAW}} ~ \sim ~ 1.2$ \textmu s; see next section). Making the assumption that propagation losses in
  free space are dominant (97\% of the energy of the resonant modes is localized
  in the free-space region of the resonator), we find that these quality factors are consistent with
  a uniform propagation loss of 70 Np/m. We note that these losses could
  also be due to beam diffraction or scattering, but propagation losses in the
  metallized regions of the resonator are expected to contribute minimally (using the values
  reported in Ref. \cite{satzingerQuantumControlSurface2018}, if these were the sole source of loss we would expect quality factors above $10^6$ for our geometry).
\end{itemize}

\begin{figure}[t]
\centering
\includegraphics{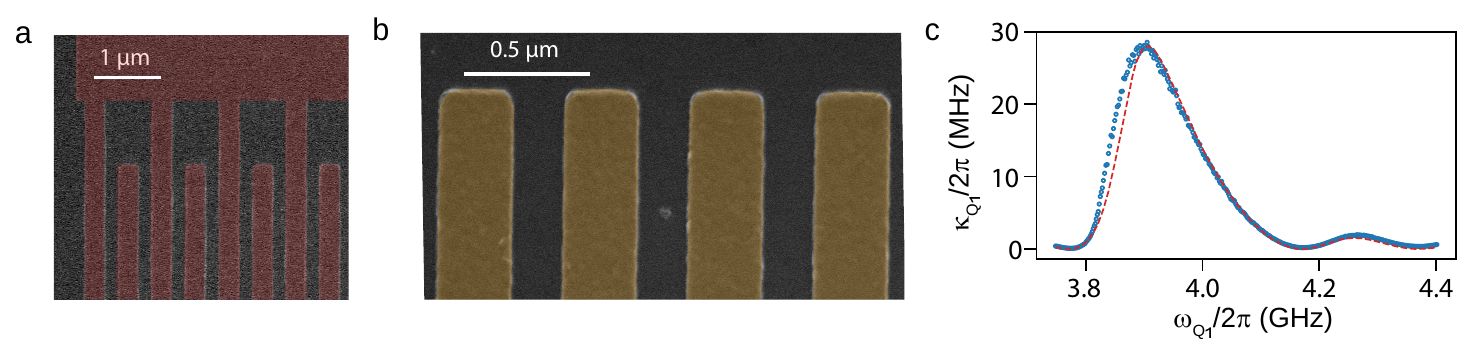}
\caption{\label{SAW}(a,b) Scanning electron micrograph of the details of the IDT and Bragg mirrors. The IDT pitch is twice the finger spacing, and is slightly smaller than twice the mirror finger spacing. (c) Extracted qubit decay rate $\kappa = 1/T_1$ from Fig. 1e, measured at maximum coupling. Decay is dominated by phonon emission from the IDT. Blue circles are extracted from an exponential decay fit, red dashed line is the prediction of a circuit model as described in the text.}
\centering
\end{figure}

\subsection{Coupling between qubit and individual SAW Fabry-Perot modes}

The qubit emission rate into the SAW phonon channel is easily determined using
qubit lifetime measurements for emission times $t$ shorter than the phonon transit time $\tau$, so that reflections, and the internal Fabry-P{\'e}rot mode structure, can be ignored. However, another useful quantity is the qubit coupling rate $g$ to the individual Fabry-P{\'e}rot acoustic modes; this measurement is not completely straightforward. This is especially true for coupling rates $g$ that are smaller than the
free spectral range $\upsilon_{\mathrm{FSR}} = 1 / \tau = 1.97$ MHz, that is, for couplings $g \tau < 1$. We
can relate the coupling rate to the qubit emission rate $\kappa$ into the SAW
channel via Fermi's golden rule: $\kappa = g^2 / \upsilon_{\mathrm{FSR}}$ \cite{milonniExponentialDecayRecurrences1983}.
This assumes that the coupling of the qubits to each individual SAW mode is identical, an approximation that seems quite reasonable given that the bandwidth of the IDT (200 MHz) is much larger than the maximum emission rate (20 MHz), so that the IDT coupling should vary little over the relevant frequency range.

The value of $g$ determined for maximum qubit-SAW coupling ($g = 2 \pi \times 2.57$ MHz)
is significantly larger than each qubit's linewidth (respectively $1 / T_{2 R} = 2 \pi \times
76$ kHz and $2 \pi \times 0.26$ MHz), larger than the resonator free spectral range
$\upsilon_{\mathrm{FSR}} = 1.97$ MHz, and larger than the SAW mode's typical loss rate $1 /
T_{1 \mathrm{SAW}} = 2 \pi \times 130$ kHz, demonstrating that, at maximum coupling, the qubit and SAW
resonator are in the strong multimode-coupling regime.

As a verification of these conclusions, we perform spectroscopy
measurements of qubit $Q_1$ as it is tuned through each of the SAW mode resonances, for different tunable coupling strengths. The results are shown in the left panels of Fig. \ref{qutip}, and vary from a series of simple avoided-level crossings at the smallest coupling,  where the individual modes are easily resolved, to a more complex dressed-states picture at larger coupling strengths.

We also measure the time-domain vacuum Rabi oscillations for the same coupler
settings, shown in the central panels of Fig. \ref{qutip}. Interestingly, no frequency dependence is observed
for $t<\tau$ before one transit time is completed (white dashed line). For $t \geqslant \tau$,  a series of collapses and revivals appear, resulting from interference from interactions with multiple
resonant modes. We note that for coupler biases $\Phi_g / \Phi_0 > 0.7$, no
frequency dependence is observed for several transits, indicating
that the qubit is now emitting itinerant phonons.

We model these results by considering a system comprising a qubit (lowering operator $\sigma_-$, and frequency $\omega_q$) coupled to a family of $n_a$ oscillators
(annihilation operator $a_j$), with frequencies $\omega_q+\Delta_j$ spaced by the resonator free spectral range
$\upsilon_{\mathrm{FSR}}$). The coupling is through a multi-mode Jaynes-Cummings Hamiltonian $H$,
\begin{equation}
    H / \hbar = \sum_{j=1}^{n_a} \Delta_j a_j^{\dag} a_j + g (\sigma_+ a_j + \sigma_- a^{\dag}_j),
\end{equation}
written in the qubit rotating frame. We do not include the second
qubit since $Q_2$'s coupler is turned off in these experiments.

We fit the qubit spectrum by solving for the eigenvalues of $H$ (black dashes on the left panels of Fig.~\ref{qutip}). Ref~\cite{milonniExponentialDecayRecurrences1983} gives an analytical formula for the  qubit excited state amplitude $A_e$ (linked to the qubit population by $P_e = |A_e|^2$) coupled to an infinite family of oscillators:
\begin{equation}\label{milonni}
    A_e =  e^{-\kappa t} \sum_{n=0}^{\infty} e^{-i n\tau (\Delta_0-i \kappa_a/2-i\kappa/2)}\times \mathcal{L}_n(\kappa(t-n\tau))\Theta(t-n\tau),\end{equation}
where $\Delta_0$ is the qubit detuning to the nearest oscillator, $\kappa_a = T_{1 \mathrm{SAW}}^{-1}$, $\Theta(t)$ is the unit step function, and the functions $ \mathcal{L}_n(t)$ are defined as:

\begin{equation}
\mathcal{L}_n(t) = \left \lbrace \begin{array}{l}
     L_0 (t), \quad n = 0\\
     L_n (t) - L_{n - 1} (t), \quad n \geqslant 1
   \end{array} \right.
\end{equation}
where $L_n$ are the Laguerre polynomials. To take qubit imperfections into account, we also  simulate the time-domain evolution by numerical integration of the Lindblad master equation \cite{lindbladGeneratorsQuantumDynamical1976,wallsQuantumOptics2008} using the python package QuTiP \cite{johanssonQuTiPOpensourcePython2012}. We include the Lindblad
collapse operators to account for qubit relaxation, $\sigma_- /
\sqrt{T_{1, \mathrm{int}}}$, qubit decoherence $\sigma_z \sqrt{\Gamma_{\phi}}$,  and oscillator
energy relaxation $a_i / \sqrt{T_{1 \mathrm{SAW}}}$ (taken to be identical for all oscillators). Here the qubit decoherence is
\begin{equation}
    \Gamma_{\phi} = \frac{1}{T_{2, R}} - \frac{1}{2 T_{1, \mathrm{int}}}.
\end{equation}
The simulations are done using $n_a = 8$ oscillator modes, each spanned by two Fock states.

The analytical results and the fits are shown respectively as red and black dashed lines in Fig. \ref{qutip}, and show good agreement with the data, supporting our determination of the coupling strength $g$. All parameters involved in the fits were arrived at independently, with the exception of the SAW energy relaxation time,  which we adjusted to $T_{1 \mathrm{SAW}} = 1.2$ \textmu s. This value is compatible with a separate measurement of the SAW resonator ($T_{1 \mathrm{SAW}} \approx 2$ \textmu s, see above).

We can link the SAW energy relaxation time to the quantum transfer efficiency $\eta$ used in the main text in the following way: Consider the initial situation where the qubit is in its excited state, and all oscillator modes are empty. Assuming a perfect transfer of energy from the qubit to an infinity of oscillator modes, energy conservation states that the occupation probabilities $\alpha_i$ of the oscillators sum to $\hbar \sum_i |\alpha_i|^2 (\omega_q+\Delta_i)=\hbar \omega_q$. During a phonon transit, each oscillator energy decays with a  timescale $T_{1 \mathrm{SAW}}$ that we assume is identical for all the oscillators. The amount of energy left in the system afterwards is $\hbar \sum_{i}|\alpha_i|^2 (\omega_q+\Delta_i) e^{- \tau / T_{1 \mathrm{SAW}}}$. This implies that, at best, the quantum transfer efficiency is $\eta \leqslant e^{- \tau / T_{1 \mathrm{SAW}}} = 0.65$. This value is in good agreement with the experiment (see main text).

\begin{figure}[t]
\centering
\includegraphics[width=\textwidth]{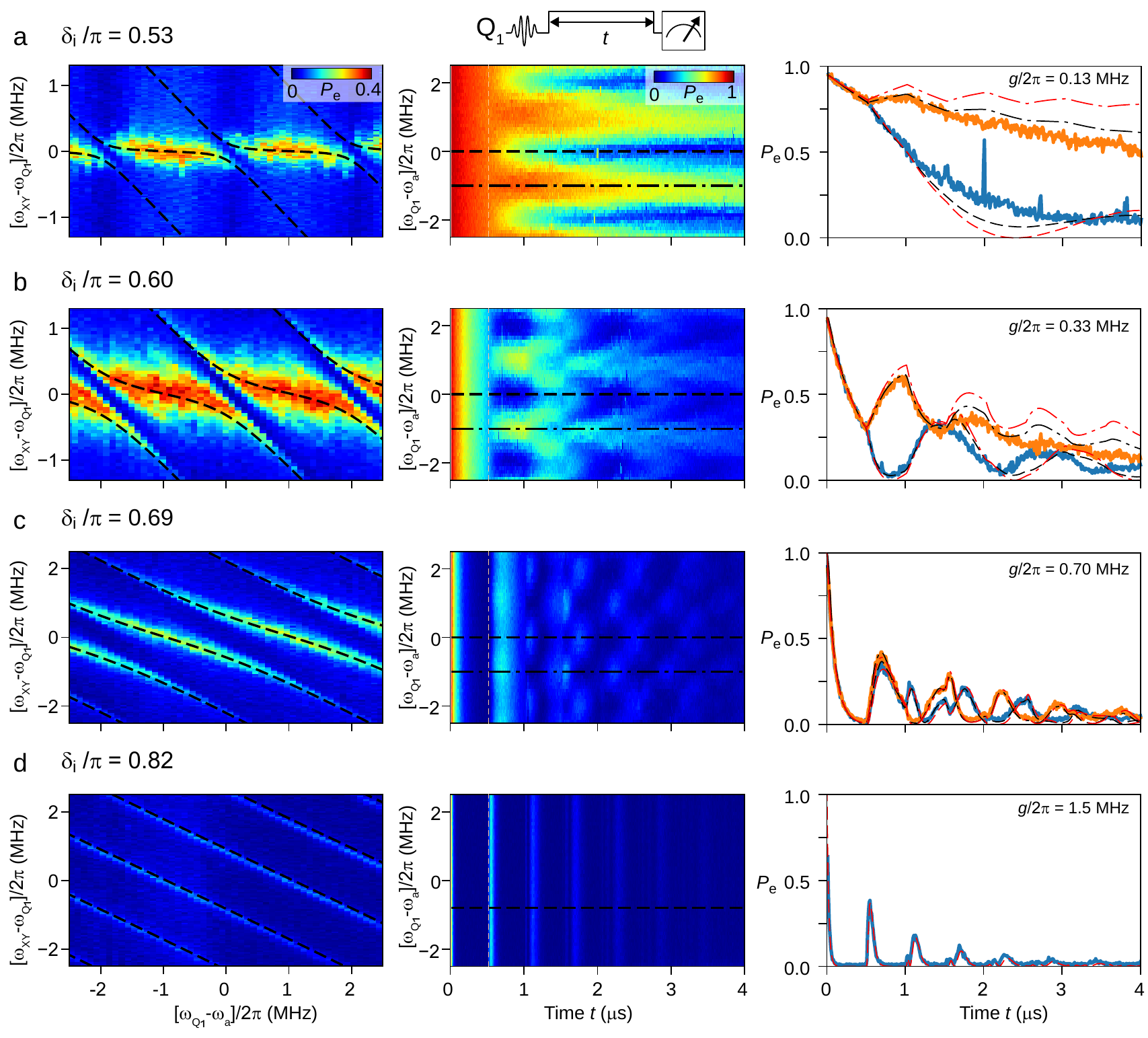}
\caption{\label{qutip} (a-d) \emph{Left}: Qubit 1 spectroscopy near the SAW Fabry-P{\'e}rot mode $\omega_a/2\pi = 3.9542$ GHz for different coupling biases corresponding to the coupling strengths indicated on each of the right panels. \emph{Center}: Corresponding energy decay rate measurements in the same frequency range. \emph{Right}: Line cuts through the data in the central panels at $\omega_q=\omega_a$ (blue curves) and $\omega_q=\omega_a-\omega_{\mathrm{FSR}}/2$ (orange curves). Black dashed lines are QuTiP simulations as described in the text, red dashed lines are from Eq.~\ref{milonni}. We did not perform a QuTiP simulation for d, as it requires the inclusion of more than 20 oscillators, which is very computationally intensive.}
\centering
\end{figure}

\section{Simulation and control pulses}

\subsection{Input-output theory}

To model the qubit emission and capture of the itinerant phonons, we use input-output theory
as derived by Gardiner and Collett \cite{gardinerInputOutputDamped1985}, modeling the qubits as oscillators. We do not include the qubit anharmonicity as we constrain the simulations to single excitations. The qubits are modeled as two intra-cavity
modes, with bosonic operators $s_1$ and $s_2$ that interact with the same
incoming and outgoing phonon fields, represented by bosonic operators $a_{\mathrm{in}}$ and
$a_{\mathrm{out}}$. In a frame rotating at the operating frequency
$\omega_0$, the operator evolution is given by
\begin{align}
\dot{s_1} = - [i \delta_1 (t) + \kappa_1 (t)] s_1 + \sqrt{\kappa_1 (t) } a_{\mathrm{in}} (t), \label{q1IO}\\
\dot{s_2} = - [i \delta_2 (t) + \kappa_2 (t)] s_2 + \sqrt{\kappa_2 (t) } a_{\mathrm{in}} (t), \label{q2IO}\\
\sqrt{\kappa_1 (t) } s_1 (t) + \sqrt{\kappa_2 (t) } s_2 (t) = a_{\mathrm{in}} (t) + a_{\mathrm{out}} (t), \label{IOeq}\\
a_{\mathrm{in}} (t) = \sqrt{\eta} a_{\mathrm{out}} (t - \tau), \label{IOwrap}
\end{align}
where $\delta_1$ and $\delta_2$ are the de-tunings of qubits $Q_1$ and $Q_2$ with respect to
$\omega_0$, $\kappa_1$ and $\kappa_2$ their respective energy relaxation rates,
and $\eta \in [0, 1]$ characterizes the finite transmission of the acoustic
channel. We note that in all the experiments presented here, either $\kappa_1$
or $\kappa_2$ is set to zero, preventing the qubits from interacting directly
with each other. Equation~(\ref{IOeq}) thus describes the interaction of a single
qubit with the incoming and outgoing phonon fields.

In all experiments described in the main text, we chose to emit wavepackets of
hyperbolic secant shape, $a_{\mathrm{out}} (t) \propto \mathrm{sech} (\kappa_c t / 2)$, where $\kappa_c$ is the characteristic bandwidth
of the wavepacket, fixed to $1/\kappa_c = 10$ ns for all
experiments. Using Eqs.~(\ref{q1IO}-\ref{IOeq}), we find that when emitting all the
energy stored in one qubit, that is, emitting one phonon, so that $s_i (t \rightarrow \infty) = 0$,
constrains $\kappa_i$ to
\begin{equation}
\kappa_i (t) = \frac{e^{\kappa_c t}}{1 + e^{\kappa_c t}},
\end{equation}
in the absence of any incoming field. However, emitting only a fraction $\alpha$ of
the qubit energy (i.e. emitting a fraction $\alpha$ of a phonon) constrains $\kappa_i$ to
\begin{equation}
\kappa_i (t) = \frac{\alpha}{1 + (1 - \alpha) e^{\kappa_c t}} \times
   \frac{e^{\kappa_c t}}{1 + e^{\kappa_c t}}.
\end{equation}
When time-reversed, these equations respectively describe the capture of a phonon
when the qubit is in its ground state, and the capture of a portion $\alpha$
of a phonon when the qubit already contains a portion ($1 - \alpha$) of a
phonon.

We use the coupler calibrations described in Section~(\ref{Couplers}) to determine the
qubit and coupler flux pulses needed to keep the qubit operating
frequency constant. We note that the flux
pulses are very nearly Gaussian of bandwidth $<20$ MHz, a shape that is ideal for our room-temperature
electronics, as these include Gaussian filters of bandwidth $250$ MHz.

We use the input-output model to simulate the time evolution of $Q_1$ in Figs. 2a, 2b, 4a and 4b of the main text, taking $\eta = 0.67$. We also use this model to simulate
the interferometry experiment shown in Fig. 3 of the main text, with a simulated efficiency $\eta = 0.67$.  The amplitude of the destructive interference, which results in qubit re-excitation and relies on the returning phonon being in phase with the phonon emitted during the capture pulse, is reduced by $1 - \eta / 2$ due to the half-phonon damping during its transit in the acoustic channel. In contrast, the amplitude of the constructive interference, which results in the qubit being completely de-excited, is not affected by acoustic losses.

To explain the non-zero qubit population for constructive interference observed in Fig. 3, one needs to take into account the de-phasing accumulated between the qubit and the itinerant phonon. It is likely that the qubit is the dominant source of de-phasing. One could phenomenologically include random noise in the qubit detuning function $\delta_1 (t)$ to model
the accumulation of a random phase $\varphi$ in the input-output model, and estimate the result using Monte-Carlo simulations. We instead choose to let the qubit be ``the clock'', i.e. take the point of view that its phase evolves without noise, and we instead model the phonon as acquiring a random phase $\varphi$. This is equivalent to defining a complex channel transmission $\eta e^{i \varphi}$. We implement the Monte-Carlo simulations by assuming the phase $\varphi$ has a Gaussian distribution with variance set by requiring $e^{- \frac{1}{2} \langle
\varphi^2 \rangle} = e^{- \tau / T_{2 R}}$ \cite{ithierDecoherenceSuperconductingQuantum2005}. The curve shown in Fig. 3 of the main text is the
average qubit population measured over 1024 realizations.

We note that including de-phasing reduces the interference visibility, as shown in Fig. 3B of the main text. A small discrepancy however remains, which may be due to pulses imperfections or a mis-calibrated de-phasing.

\subsubsection{Cascaded quantum systems}

To simulate the impacts of acoustic channel loss and qubit de-phasing on the
fidelity of the quantum state transfers described in Figs. 2 and 4  of the main text, and of the Bell state creation described in Fig. 4 of the main text, we adapt the quantum cascaded systems model
described in Refs. \cite{gardinerDrivingQuantumSystem1993, carmichaelQuantumTrajectoryTheory1993}. This model describes the evolution of
an emitter-receiver system, where the emitter system, with density matrix
$\rho_E$, emits photons or other quanta, to which the receiver system, with density matrix
$\rho_R$, reacts.

In our experiment, labeling each qubit as emitter or as receiver is not straightforward: In  the single qubit reflectometry experiment of Figs. 2a and b, qubit $Q_1$ is both emitter and receiver, and for the double state transfer of Fig. 4a, both qubits are in turn emitter and then receiver for the other qubit.

However, the system can still be modeled using a time-delayed cascaded system model. To understand this, let us first focus on the single qubit reflectometry experiment. Qubit $Q_1$ is both emitter and receiver, becoming a receiver at a time-delayed $\tau$: We can say that $Q_1$ evaluated at time $t$ is the receiver of $Q_1$ evaluated at time $t-\tau$. More generally, the system  $\rho_R (t) = \rho_1(t) \otimes \rho_2(t)$, comprising both qubits evaluated at time $t$, is the receiver of the phonons emitted by the system at time $t-\tau$, $\rho_E (t) = \rho_1(t-\tau) \otimes \rho_2(t-\tau)$.

This is close to what is described in the time-delayed quantum feedback theory of Refs.~\cite{grimsmoTimeDelayedQuantumFeedback2015, pichlerPhotonicCircuitsTime2016,whalenOPENQUANTUMSYSTEMS2015}. In these models, the system (here, the two qubits) are interacting with a channel (here, the acoustic channel formed by the SAW resonator) at times $t$ and $t-\tau$. The difference between these models and our system lies in that the qubits can interact with the channel at any integer multiple of the phonon transit time, whereas in \cite{grimsmoTimeDelayedQuantumFeedback2015, pichlerPhotonicCircuitsTime2016}, the number of interactions is limited to two. Ref. \cite{whalenOPENQUANTUMSYSTEMS2015} however demonstrated that the time-delayed quantum feedback theory is accurate if one only considers the first two interactions, i.e. when only considering a single phonon transit through the resonator. We wish to apply these models to our experiments. For this purpose, we need to analyze the system evolution in two different situations.

\vspace{0.5cm}

First, for times $0 \leqslant t \leqslant \tau$, all emitted quanta are propagating through the channel, and no feedback has occurred. The system evolution is thus simply described by  $\rho_R(t) = \rho_1(t) \otimes \rho_2(t)$. Its evolution is given by
\begin{equation}\label{integration1}
    \dot{\rho}_R (t) =\mathcal{L} [\rho_R (t)],
\end{equation}
where the Liouvillian $\mathcal{L}$ is simply given by $i [\rho_R,H_0(t)]+ \mathcal{D} \left[ \sqrt{\kappa_1(t)} s_1 \right] \rho_R + \mathcal{D} \left[ \sqrt{\kappa_2(t)} s_2 \right]\rho_R$ with
\begin{equation}
    H_0(t) = \delta_1 (t) s_{1}^{\dag}(t) s_{1}(t) + \delta_2 (t) s_{2}^{\dag}(t) s_{2}(t),\end{equation}
where $\mathcal{D}$ is the Lindblad damping super-operator, defined by $\mathcal{D} [X]\rho = X \rho X^{\dag} - \frac{1}{2} \left \lbrace X^{\dag} X, \rho \right \rbrace$, describing the emission of $Q_1$ and $Q_2$ into the channel with their respective couplings $\kappa_1(t)$ and $\kappa_2(t)$.

\vspace{0.5cm}
Second, for times $\tau \leqslant t \leqslant 2\tau$, the quanta emitted during times $0 \leqslant t \leqslant \tau$ now interact with the system. Following Ref.~\cite{whalenOPENQUANTUMSYSTEMS2015}, we thus need to use quantum cascade theory with an imaginary  system comprising two copies of the actual physical system, evaluated at time $t$ and $t-\tau$, with density operator $\rho_S = \rho_E \otimes \rho_R$. Explicitly, this imaginary system contains four qubits,
\begin{equation}
   \rho_S (t) = \rho_1(t) \otimes \rho_2(t) \otimes \rho_1(t-\tau) \otimes \rho_2(t-\tau),
\end{equation}
each described by its own bosonic operator:
\begin{align}
    \nonumber	s_{1,R} &= s_1(t),\\
    \nonumber	s_{2,R} &= s_2(t),\\
    \nonumber	s_{1,E} &= s_1(t-\tau),\\
    \nonumber	s_{2,E} &= s_2(t-\tau).
\end{align}
The system evolution in the frame rotating at the operating frequency $\omega_0$  is given by:
\begin{equation}\label{integration2}
    \dot{\rho}_S (t) = \mathcal{L}_2 [\rho_S (t)],
\end{equation}
where $\mathcal{L}_2$ is given by \cite{whalenOPENQUANTUMSYSTEMS2015,gardinerQuantumNoiseHandbook2004}:
\begin{align}\label{L2}
   \mathcal{L}_2 = i [\rho_S, H_0 (t) + H_{\mathrm{int}} (t)] &+
   \sqrt{\eta} \mathcal{D} \left[ \sqrt{\kappa_E(t)} s_E + \sqrt{\kappa_R(t)} s_R
   \right] \rho_S  \nonumber \\ & + \left( 1 - \sqrt{\eta} \right) \mathcal{D} \left[
   \sqrt{\kappa_E(t)} s_E \right] \rho_S  \nonumber \\ &+ \left( 1 - \sqrt{\eta} \right)
   \mathcal{D} \left[ \sqrt{\kappa_R(t)} s_R \right] \rho_S,
\end{align}
with
\begin{align}
    H_{\mathrm{int}} &= \frac{i}{2} \sqrt{\eta \kappa_E (t) \kappa_R (t)} \left (s_E^{\dag} s_R - s_E s_R^{\dag} \right ).
\end{align}
In Eq.~\ref{L2}, the Lindblad super-operators and the Hamiltonian $H_{\mathrm{int}}$ describe how the output of the emitting copy $E$ is cascaded to the input of the receiving copy \cite{gardinerQuantumNoiseHandbook2004}. Since in all our experiments, only one qubit interacts with the SAW resonator at any given time, we have directly simplified the expression for the Liouvillian so as to include only terms involving one emitter qubit, $s_E = s_{1, E}$ or $s_{2, E}$, and one receiver qubit, $s_R = s_{1, R}$ or $s_{2, R}$, depending on which emitting coupling rates, $\kappa_E (t) = \kappa_1 (t - \tau)$ or $\kappa_2 (t - \tau)$, and which receiving coupling rates, $\kappa_R (t) = \kappa_1(t)$ or $\kappa_2 (t)$, are non-zero.

The Hamiltonian $H_{\mathrm{0}}$ describes the evolution of the system in the absence of any
interaction with the acoustic channel. It is given by:
\begin{align}
    H_{\mathrm{0}} =& \delta_1 (t) s_{1,R}^{\dag}(t) s_{1,R}(t) \\\nonumber +& \delta_2 (t) s_{2,R}^{\dag}(t) s_{2,R}(t)
    \\\nonumber +&\delta_1 (t-\tau) s_{1,R}^{\dag}(t) s_{1,R}(t) \\\nonumber+& \delta_2 (t-\tau) s_{2,R}^{\dag}(t) s_{2,R}(t).
\end{align}

\vspace{0.5cm}
To obtain the system evolution, we first integrate Eq.~\ref{integration1} with initial conditions $\rho_R (0) = \rho_1(0) \otimes \rho_2(0)$ for times $0 \leqslant t \leqslant \tau$. We then integrate Eq.~\ref{integration2} for times $\tau \leqslant t \leqslant 2\tau$ with initial condition $\rho_E (\tau) =  \rho_1(0) \otimes \rho_2(0)$ and $\rho_R (\tau^+) = \rho_R (\tau^-)$, that is to say, with the final state reached during the first integration.

\vspace{0.5cm}
To model the experiments of Figs. 2 and 4, we use the control pulses $\kappa_1 (t)$ and $\kappa_2 (t)$ shown in the insets as input and we take $\delta_1 (t)=0$ and $\delta_2 (t)=0$. We set the transmission channel efficiency to be $\eta = 0.67$. We also include qubit imperfections by including the collapse operators for qubit de-phasing $s_{i, j} / \sqrt{T_{1, \mathrm{int}, i}}$, and qubit decoherence $s_{i, j}^{\dag}
s_{i, j} \sqrt{\Gamma_{\phi, i}}$, with
\begin{equation}
    \Gamma_{\phi, i} = \frac{1}{T_{2, R, i}} - \frac{1}{2 T_{1, \mathrm{int}, i}},
\end{equation}
using the parameter values as given in Table S1. The models, and in particular the modeled density matrices, show very good agreement to the data.

\section{Additional quantum state transfer measurements}
In Table~\ref{table:fids} we show some additional measurements for characterizing the swap operations, in terms of their fidelity and trace distance.
\begin{table}[h!]
\centering
\begin{tabular}{lccc}
  \hline
  Swap operation & Fidelity & Fidelity with readout correction & Trace distance\\
  & $\mathrm{Tr} \left( \chi_{\mathrm{m}} \cdot \chi_{\mathrm{ideal}} \right)$&
   $\mathrm{Tr} \left( \chi_{\mathrm{m}}^{\ast} \cdot \chi_{\mathrm{ideal}} \right)$ &
  $\sqrt{\mathrm{Tr} \left( \left[ \chi_{\mathrm{m}} - \chi_{\mathrm{sim}}
  \right]^2 \right)}$\\
  \hline
  
  \textbf{1-phonon swaps} & & & \\
  $Q_1 \rightarrow Q_1$ (see Fig. 2a) & $0.83 \pm 0.002$& $0.84 \pm 0.002$ & 0.04\\
  $Q_1 \rightarrow Q_2$ & $0.81 \pm 0.002$& $0.82 \pm 0.002$ & 0.07\\
  $Q_2 \rightarrow Q_2$ & $0.77 \pm 0.002$ & $0.78 \pm 0.002$ & 0.06\\
  $Q_2 \rightarrow Q_1$ & $0.71 \pm 0.002$ & $0.72 \pm 0.002$& 0.1\\
  \textbf{2-phonon swap} (see Fig. 4a) & & & \\
  $\left \lbrace \begin{array}{cccc}
    Q_1 & \rightarrow & Q_2 & \\
    & Q_2 & \rightarrow & Q_1
  \end{array} \right.$ & $0.63 \pm 0.01 $ & $0.64 \pm 0.01 $  & $0.54$ \\
  \hline
\end{tabular}
\caption{\label{table:fids}Measured fidelities and trace distances for all quantum state transfer configurations.}
\end{table}

\end{document}